\documentstyle[prd,aps,epsf,floats]{revtex} 
\draft

\begin{document}
\twocolumn[\hsize\textwidth\columnwidth\hsize\csname
@twocolumnfalse\endcsname
\title{
\hbox to\hsize{\large Submitted to Phys.~Rev.~D \hfil E-Print
astro-ph/9708420}
\vskip1.55cm
Ultra-high energy neutrino-nucleon cross section and radiative
corrections}
\author{G\"unter Sigl}
\address{Department of Astronomy and Astrophysics, The
University of Chicago, Chicago, Illinois 60637-1433}
\date{\today}
\maketitle
\begin{abstract}
A cubic kilometer scale experiment has been proposed to detect
cosmic neutrinos of energy from tens of GeV up to the highest
energies observed for cosmic rays, $\sim10^{20}\,$eV, or
possibly even beyond. Detection efficiencies depend
crucially on the neutrino-nucleon cross section at these
energies at which radiative corrections beyond the lowest order
approximation could become non-negligible. The differential cross
sections can be modified by more than $50\%$ in some regions of
phase space. Estimates of corrections to the quantities most
relevant for neutrino detection at these energies give, however,
less dramatic effects: The average inelasticity in the outgoing lepton
is increased from $\simeq0.19$ to $\simeq0.24$. The inclusive cross
section is reduced by roughly half a percent. The dominant uncertainty of
the standard model ultra-high energy neutrino-nucleon cross section
therefore still comes from uncertainties of the parton distributions in
the nucleon at very low momentum fractions.
\end{abstract}
\vskip2.2pc]

%%%%%%%%%%%%%%%%%%%%%%%%%%%%%%%%%%%%%%%%%%%%%%%%%%%%%%%%%%%%%%%%%%%%%%
%% Main Text %%%%%%%%%%%%%%%%%%%%%%%%%%%%%%%%%%%%%%%%%%%%%%%%%%%%%%%%%
%%%%%%%%%%%%%%%%%%%%%%%%%%%%%%%%%%%%%%%%%%%%%%%%%%%%%%%%%%%%%%%%%%%%%%

\narrowtext

\section{Introduction}
Several proposals have recently been put forward for the search
for cosmic neutrinos above tens of GeV up to the high energy end
of the cosmic ray spectrum, and possibly beyond. The most well
developed technique is to detect Cherenkov light from the muon
produced in a charged-current reaction of an ultra-high energy
(UHE) muon neutrino with a nucleon in either ice or
water. Several prototype detectors based on this technique have been
constructed, namely DUMAND at a depth of nearly $5\,$km in the
ocean near Hawaii (now out of commission), the neutrino telescope
at Lake Baikal about $1\,$km deep, NESTOR at about $3.5\,$km
depth in the Mediterranean near Greece, and AMANDA up to $2\,$km
deep in South Pole ice~\cite{ghs}. Other techniques have been
proposed such as detection of horizontal air
showers~\cite{ydjs,bpvz}, or detection of
acoustic~\cite{acoustic} or radio waves~\cite{radio} associated
with the neutrino induced cascades.

For a given neutrino flux, the efficiency of all these methods
depends predominantly on the neutrino-nucleon interaction
cross section. To lowest order in the electroweak (EW) coupling,
this cross section has been discussed in detail in the
literature, see, e.g., Refs.~\cite{fkr,gqrs} for the most recent
work. Due to uncertainties in the extrapolation of quark
distribution functions in the nucleon to very small fractional
momentum transfers, $x\lesssim10^{-4}$, and large (negative)
squared four-momentum transfers, $Q^2\gtrsim10^5\,{\rm GeV}^2$,
best estimates for energies around $10^{20}\,$eV (in the
laboratory frame) vary by factors of a few.

On the other hand it is well known that higher order processes
can become important or even dominant for electromagnetic (EM)
interactions at UHE. For example, energy exchange between two
leptons $l_1$ and $l_2$ is dominated by EM bremsstrahlung,
$l_1+l_2\rightarrow l_1+l_2+\gamma$, for center-of-mass energies
$s$ exceeding the square of the electron mass, rather than by
ordinary Mott scattering, $l_1+l_2\rightarrow l_1+l_2$. The
relevant cross section rises with the logarithm of $s$.

EW radiative corrections to deep inelastic
neutrino-nucleon scattering have been calculated before in the
literature: Ref.~\cite{rpsn} contains a discussion of the
leading log approximation for which only corrections involving
photons are relevant. These corrections contain a factor
$\ln(s/m_l^2)$ where $m_l$ is the charged lepton mass and also
depend on the behavior of the parton distribution functions.
Single and double differential cross sections have been
evaluated in Ref.~\cite{rpsn} for neutrino energies $E_\nu$ up to a few
hundred GeV, using rough estimates of the parton distributions
available back then.

To compare with HERA measurements~\cite{hera} at $s\simeq10^5\,{\rm
GeV}^2$, corresponding to $E_\nu\simeq50\,$TeV, more recently
complete analytical expressions have been presented in
Refs.~\cite{bbcr,spiesberger}. At these energies, corrections
have been shown to be up to $50\%$ in certain areas of phase
space. Consequently, at energies approaching $10^{20}\,$eV,
radiative corrections could be larger still and may play an
important role. We therefore found it worthwhile to extend
estimates of radiative corrections to UHE, using updated parton
distribution functions and with a special emphasize on the
quantities relevant for UHE neutrino detection.

The rest of this paper is organized as follows: In Sec.~II we
introduce our estimates of radiative corrections at UHE. In
Sec.~III we present numerical results. We finally summarize our
findings and resulting consequences in Sec.~IV.

\section{Estimates of radiative corrections at ultra-high
energies}

In principle, the full EW radiative corrections complete to
order $g^2$ where $g$ is the EW coupling constant can be
computed from the expressions given in Ref.~\cite{bbcr} and
references therein. These, however, involve hundreds of
terms and are somewhat hard to reproduce. In contrast, the
leading logarithmic approximation is comparatively simple and
is expected to be very good at the UHEs we are interested
in~\cite{rpsn}. Furthermore, to be concrete we will restrict
ourselves to
charged-current reactions which are more relevant for most of
the detection methods relying on Cherenkov radiation from
muons. We use the usual kinematic variables, $Q^2=2ME_\nu xy$,
with $M$ the nucleon mass, $y=(E_\nu-E_l)/E_\nu$, and $E_l$ the
energy of the outgoing charged lepton in the
laboratory frame. The contribution from the lepton leg to the
double differential cross section in
leading log approximation can then be written as
\begin{eqnarray}
  \frac{d^2\sigma}{dxdy}&=&\frac{d^2\sigma_0}{dxdy}+
  \frac{\alpha}{2\pi}\ln\frac{2ME_\nu(1-y+xy)^2}{m_l^2}
  \label{ll}\\
  &&\times\int_0^1dz\frac{1+z^2}{1-z}\biggl[
  \frac{y\,\Theta(z-z_{\rm min})}{z(y+z-1)}
  \frac{d^2\sigma_0}{dxdy}(\tilde{x},\tilde{y})\nonumber\\
  &&\hskip2.5cm-\frac{d^2\sigma_0}{dxdy}(x,y)\biggr]\nonumber\,,
\end{eqnarray}
where $\alpha\simeq1/137$ is the EM fine structure constant,
\begin{eqnarray}
  \tilde{x}&=&\frac{xy}{z+y-1}\nonumber\\
  \tilde{y}&=&\frac{z+y-1}{z}\label{xyz}\\
  z_{\rm min}&=&1-y+xy\nonumber\,,
\end{eqnarray}
and $d^2\sigma_0/dxdy$ is the lowest order cross
section. Eq.~(\ref{ll}) is expected to be a good approximation
for $\ln(2ME_nu/m_l^2)\gg\ln(1-y)$~\cite{rpsn}. Assuming
isoscalar nucleons (i.e., averaging over
protons and neutrons), the average of the lowest order cross
section over neutrinos and antineutrinos is given by
\begin{eqnarray}
  \frac{d^2\sigma_0}{dxdy}&=&\frac{2G_{\rm F}^2ME_\nu x}{\pi}
  \left(\frac{m_W^2}{Q^2+m_W^2}\right)^2(2-2y+y^2)\nonumber\\
  &&\times\left[\frac{1}{2}q_v(Q^2,x)+q_s(Q^2,x)\right]
  \,.\label{sigma0}
\end{eqnarray}
Here, $G_{\rm F}$ is Fermi's constant, $m_W$ is the mass of the
$W$ boson, and $q_v(Q^2,x)$ and $q_s(Q^2,x)$ are the valence and
sea quark distributions of the nucleon.

The first term under the integral in Eq.~(\ref{ll}) comes from
bremsstrahlung of the lepton. Its $z\to1$ infrared divergence is
cancelled by the second term under the integral which describes
lepton propagator corrections due to virtual photons such that
the total result is finite. It was argued in Ref.~\cite{rpsn}
that the lepton contribution to the QED correction is
dominant and that contributions from the quarks do not have to
be included because they are accounted for in the quark
distribution functions.

For comparison we also calculated the hard photon bremsstrahlung
contribution $\sigma_{\rm hb}(E_\nu)$ to the inclusive
cross section. In the following we denote the modulus of the
3-momenta of the outgoing photon and the quark by $k$ and
$q^\prime$, respectively, and we also use $q^\prime_0=((q^\prime)^2+
M^2)^{1/2}$. Furthermore, let $p$ be the modulus of the 3-momentum
of the incoming neutrino and $\mu_{kp}$, $\mu_{kq^\prime}$, and
$\mu_{pq^\prime}$ the cosine of the angle between the 3-momenta
indicated. After integrating out the $\delta-$functions
for 4-momentum conservation we have
\begin{eqnarray}
  \sigma_{\rm hb}(E_\nu)&=&\frac{1}{4}\frac{1}{(2\pi)^4}
  \frac{Mm_l}{E_\nu^2}\int dkdq^\prime d\mu_{kp}
  d\mu_{kq^\prime}dx\nonumber\\
  &&\times\frac{k}{\left[(1-\mu_{kp}^2)(1-\mu_{kq^\prime}^2)-
  (\mu_{pq^\prime}-\mu_{kp}\mu_{kq^\prime})^2\right]^{1/2}}
  \nonumber\\
  &&\times\,\overline{{\cal M}^2}\,\frac{q_v(Q^2,x)+q_s(Q^2,x)}{x^2}
  \,,\label{sigmabrems}
\end{eqnarray}
where $\mu_{pq^\prime}$ is given by
\begin{eqnarray}
  \mu_{pq^\prime}&=&\frac{q^\prime_0}{q^\prime}
  \left(1+\frac{M}{xE_\nu}\right)+\frac{k}{xE_\nu}
  \left(\mu_{kq^\prime}-\frac{q^\prime_0}{q^\prime}\right)
  \nonumber\\
  &&+\frac{k}{q^\prime}\left(1-\mu_{kp}\right)
  -\frac{M}{q^\prime}+
  \frac{M(k-M)+m_l^2/2}{q^\prime xE_\nu}
  \,,\label{mupq1}
\end{eqnarray}
and the integration in Eq.~(\ref{sigmabrems}) is performed over
all areas where $-1\leq\mu_{pq^\prime}\leq1$ and where the angle
dependent factor in the integrand is real. The squared matrix
element $\overline{{\cal M}^2}$ in Eq.~(\ref{sigmabrems}) is averaged
and summed over polarizations of incoming and outgoing particles,
respectively, as well as over the electric charges of the quarks
in the relevant distributions. In computing this matrix element
we used the software REDUCE. The Feynman diagrams for bremsstrahlung
photons attached to the incoming and outgoing quark, the outgoing
lepton and to the intermediate $W$ boson as well as all interference
terms were included. We note that as opposed to Eq.~(\ref{ll}) this
causes some double counting of radiative quark corrections, when using
the same quark distribution functions as for Eq.~(\ref{ll}). This is,
however, expected to cause only a small error~\cite{spiesberger}.

As expected, Eq.~(\ref{sigmabrems}) diverges logarithmically for
$k\to0$. One can, however, still get a rough estimate of the
radiative corrections without having to compute the virtual
corrections. To this end one observes that a natural cutoff scale
for $k$ is given by the quark and lepton masses, and we will
therefore restrict the range of integration to
$k\geq{\rm min}(M,m_l)$. This will not provide a precise value
for the EM radiative corrections but at least a rough estimate.

\begin{figure}
\centering\leavevmode
\epsfxsize=3.2in
\epsfysize=3.2in
\epsfbox{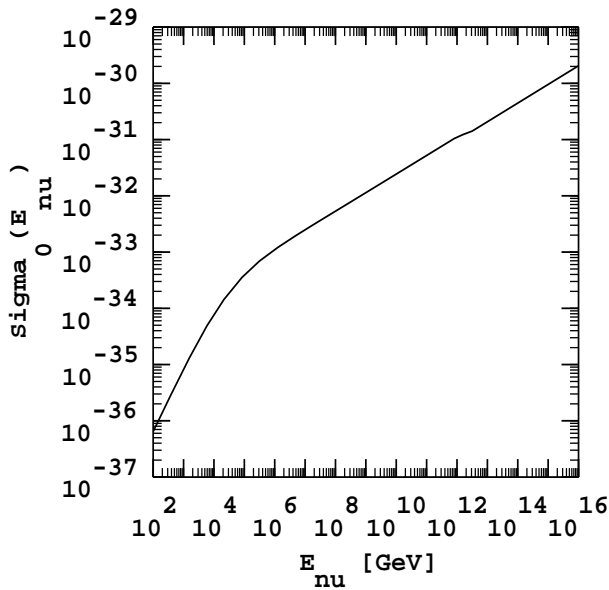}
\caption[...]{The lowest order inclusive charged-current
neutrino-nucleon cross section as given by Eqs.~(\ref{sigma0})
and~(\ref{qquark}) as a function of neutrino energy in the
nucleon rest frame.}
\label{F1}
\end{figure}

For the quark distribution functions we use the CTEQ3 distributions
in the deep-inelastic scattering factorization scheme
parametrization at $Q^2=(1.6\,{\rm GeV})^2$~\cite{cteq}. Assuming a
scaling of the sea quark distribution with $\ln Q^2$, we adopt
\begin{eqnarray}
  xq_v(Q^2,x)&=&\frac{1}{2}\biggl[1.36x^{0.47}(1-x)^{3.51}
  (1+6.19x^{1.04})\nonumber\\
  &&\hskip0.5cm+0.837x^{0.47}(1-x)^{4.22}(1+2.58x^{0.748})\biggr]
  \nonumber\\
  xq_s(Q^2,x)&=&max\left[\frac{\ln(Q^2/{\rm GeV}^2)}
  {\ln1.6^2},1\right]\label{qquark}\\
  &&\times\left[0.033x^{-0.332}(1-x)^{8.16}(1+23.2x)\right]
  \,.\nonumber
\end{eqnarray}
This analytical form yields good agreement of the lowest order
cross sections with Ref.~\cite{gqrs} (see Fig.~\ref{F1}).
Although these distributions are uncertain within up to factors
of a few for $x\ll10^{-4}$ and $Q^2\gg m_W^2$,
our results for the fractional radiative corrections are relatively
insensitive to these uncertainties.

The cross sections are also practically
independent of the lepton mass at these energies and therefore apply
to all neutrino flavors although the resulting shower development
is different for electrons and tau leptons.

\section{Numerical results}

We now present results of numerical evaluations of
Eqs.~(\ref{ll}) and (\ref{sigmabrems}). We first note that for
neutrino detection at UHE the probably
most relevant quantity is the inclusive charged-current cross
section $\sigma(E_\nu)$ because, unlike in an accelerator
experiment such as HERA, the proposed methods are not very
sensitive to the phase space distribution of produced hadrons
and charged leptons but rather act as calorimetric detection
techniques. The only exception may be the quantity $y$ which
is the fraction of the neutrino energy initially going into
the hadronic channel of the cascade ensuing, and equals 1 minus
the fraction transferred to the outgoing charged lepton whose
Cherenkov radiation is observed in the optical techniques.
We therefore restrict ourselves to the quantities $\sigma(E_\nu)$
and $y(d\sigma/dy)(E_\nu,y)/\sigma(E_\nu)$.

\begin{figure}
\centering\leavevmode
\epsfxsize=3.2in
\epsfysize=3.2in
\epsfbox{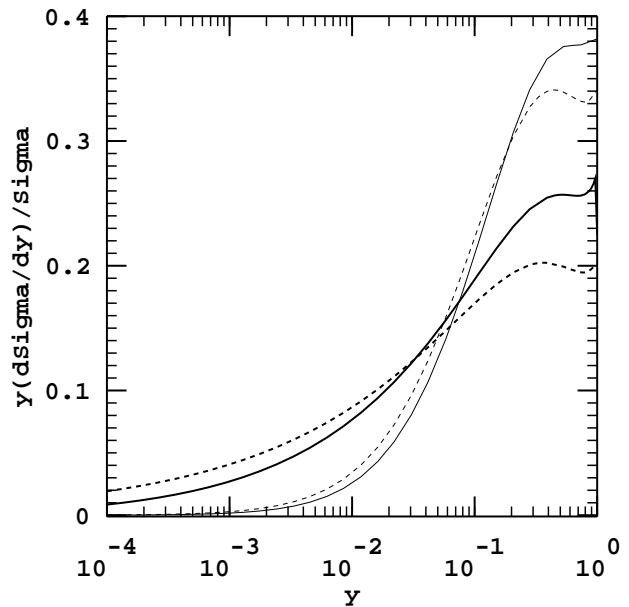}
\caption[...]{The logarithmic distribution of $y=1-E_l/E_\nu$ from the
lowest order calculation, Eq.~(\ref{sigma0}) (dashed lines) and
including radiative corrections calculated in leading
logarithmic approximation, Eq.~(\ref{ll}) (solid lines) for
$E_\nu=10^{20}\,$eV (thick lines) and for $E_\nu=10^5\,$eV (thin
lines).}
\label{F2}
\end{figure}

\begin{figure}
\centering\leavevmode
\epsfxsize=3.2in
\epsfysize=3.2in
\epsfbox{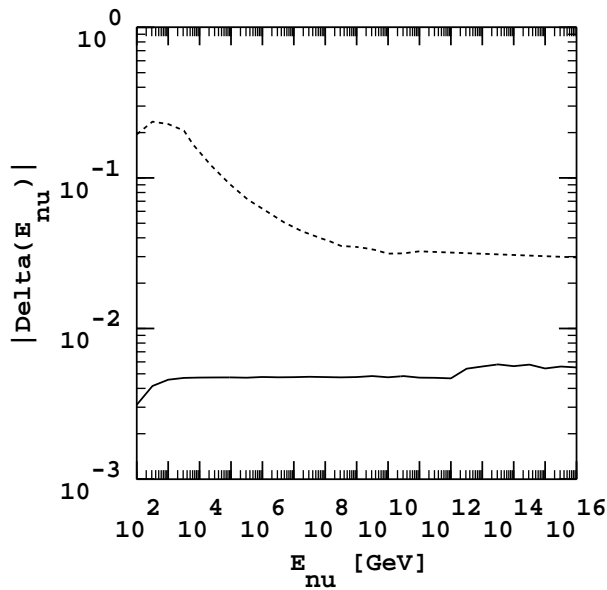}
\caption[...]{Absolute values of fractional radiative corrections
$\delta(E_\nu)=\sigma(E_\nu)/\sigma_0(E_\nu)-1$ to the
inclusive lowest order cross section as functions of
$E_\nu$. The solid line shows the negative of the
total correction in leading log approximation [Eq.~(\ref{ll})]
and  the dashed line is the hard bremsstrahlung contribution
[Eq.~(\ref{sigmabrems})]. Wiggles in the curves are due to the
finite accuracy of the Monte Carlo integration which is at the
few percent level.}
\label{F3}
\end{figure}

Fig.~\ref{F2} presents the distribution of $y$, comparing the
lowest order result from Eq.~(\ref{sigma0}) and the result
including radiative corrections in leading log approximation
from Eq.~(\ref{ll}) for a
neutrino energy $E_\nu=10^{20}\,$eV and $E_\nu=10^{5}\,$eV. The
resulting values for the average
inelasticity $\left\langle y\right\rangle$ are 0.190 and 0.307 for
the lowest order cross section, and 0.240 and 0.334 for the
corrected one, respectively. We note in passing that
corresponding values for the
neutral-current neutrino-nucleon scattering cross sections are
similar. The up to 3-dimensional integrals
involved in these calculations have
been evaluated using standard numerical
techniques.

Fig.~\ref{F3} compares the fractional corrections
$\delta(E_\nu)\equiv\sigma(E_\nu)/\sigma_0(E_\nu)-1$ to the
inclusive lowest order cross section resulting from
Eqs.~(\ref{ll}) and (\ref{sigmabrems}) as functions of $E_\nu$.
To perform the 3- and 5-dimensional integrations, respectively,
we used the adaptive Monte Carlo routine VEGAS developed by
Peter Lepage~\cite{lepage} in the version from
Ref.~\cite{press}. In order to ensure a reasonably smooth
integrand and sufficiently accurate results, we resorted to
integration variables that are power laws in $x$ and, typically,
logarithms in the other variables. Note that the leading log
approximation for the total correction is negative and roughly
an order of magnitude smaller than the hard bremsstrahlung
contribution which tends to be cancelled by virtual corrections.

\section{Discussion and conclusions}

As can be seen from Fig.~\ref{F2}, at energies around
$10^{20}\,$eV the radiative
corrections to the single differential charged-current
neutrino-nucleon cross section $(d\sigma/dy)(E_\nu,y)$ are negative
for $y\lesssim0.02$ and positive otherwise. For
$y\lesssim10^{-3}$ they grow larger than 50\%, whereas for
$y\gtrsim0.2$ they are of the order of 30\%. The average
inelasticity $\left\langle y\right\rangle$ is increased from
$\simeq0.19$ to $\simeq0.24$ whose potential influence on the
development of the neutrino induced cascade is probably the
strongest effect of radiative corrections. The corrections to
the total cross section are negative and roughly constant
at about half a percent (see Fig.~\ref{F3}). Apart from physics
beyond the standard model (see, e.g., Refs~\cite{bkklz,bhfpt},
uncertainties of the UHE neutrino-nucleon cross
section to date therefore are by an ample margin dominated by
the uncertainties in
the parton distributions in the nucleon at very low momentum
fractions.

\section*{Acknowledgments}
This work was supported, in part,
by the DoE, NSF, and NASA at the University of Chicago, and
by the DoE and by NASA through grant NAG 5-2788 at Fermilab.

%%%%%%%%%%%%%%%%%%%%%%%%%%%%%%%%%%%%%%%%%%%%%%%%%%%%%%%%%%%%%%%%%%%%%%
%% References %%%%%%%%%%%%%%%%%%%%%%%%%%%%%%%%%%%%%%%%%%%%%%%%%%%%%%%%
%%%%%%%%%%%%%%%%%%%%%%%%%%%%%%%%%%%%%%%%%%%%%%%%%%%%%%%%%%%%%%%%%%%%%%

\end{document}